\documentclass{aastex}
\usepackage{emulateapj5}

\newcommand{\Msun}{\mbox{ M}_{\odot}}
\newcommand{\gae}{\mathrel{>\kern-1.0em\lower0.9ex\hbox{$\sim$}}}
\newcommand{\lae}{\mathrel{<\kern-1.0em\lower0.9ex\hbox{$\sim$}}}
\newcommand{\kms}{km~s$^{-1}$}

\lefthead{Boroson et al.}
\righthead{QPOs in Her~X-1}

\slugcomment{Accepted for publication in the Astrophysical Journal, 2000,
v545}
\begin{document}

\title{Discovery of mHz UV Quasiperiodic Oscillations in
Hercules~X-1}

\author{Bram Boroson\altaffilmark{1}}
\affil{Goddard Space Flight Center, Greenbelt, MD 20771;
bboroson@falafel.gsfc.nasa.gov}

\and

\author{Kieran O'Brien, Keith Horne}
\affil{Physics and Astronomy, University of St. Andrews, North Haugh, St. 
Anderws, Fife KY 16 988, UK;
kso@st-andrews.ac.uk, kdh1@st-and.ac.uk}

\and

\author{Timothy Kallman, Martin
Still\altaffilmark{2}, and Patricia T. Boyd\altaffilmark{2}}
\affil{
Goddard Space Flight Center;
tim@xstar.gsfc.nasa.gov,
still@chunky.gsfc.nasa.gov,
padi@dragons.gsfc.nasa.gov}

\and

\author{Hannah Quaintrell}
\affil{The Open University, Walton Hall, Milton Keynes MK7 6AA,
UK; H.Quaintrell@open.ac.uk}

\and

\author{Saeqa Dil Vrtilek}
\affil{Harvard-Smithsonian Center for
Astrophysics, 60 Garden Street, Cambridge, MA 02138;
svrtilek@cfa.harvard.edu}

\altaffiltext{1}{National Research Council Associate}
\altaffiltext{2}{Universities Space Research Association}


\begin{abstract}

Observations of the ultraviolet continuum of the X-ray binary
system Her~X-1/HZ~Herculis with the Space Telescope Imaging Spectrograph
on the Hubble Space Telescope show quasiperiodic
oscillations (QPOs) at frequencies of 8$\pm2$ and 43$\pm2$ mHz, with rms
amplitudes of
2 and 4\%\ of the steady flux.  Observations with the Keck telescope
confirm the presence of the higher frequency QPO in the optical
continuum, with a rms amplitude of $1.6\pm0.2$\%.
The QPOs are most prominent in the HST data near
$\phi=0.5$ (where $\phi=0$ is the middle of the X-ray eclipse), suggesting
that they arise not in the accretion disk but on the X-ray heated face of
the companion star.  We discuss scenarios in which the companion star
reprocesses oscillations in the disk which are caused by either Keplerian
rotation or a beat frequency between the neutron star spin and Keplerian
rotation at some radius in the accretion disk.

\end{abstract}

\section{Introduction}

Hercules~X-1/HZ~Herculis is an X-ray binary consisting of a 1.24 second
pulsar in an eclipsing 1.7 day orbit with a $\sim2\Msun$ mass normal
companion. As a result of its many periodicities, it is one of the most
frequently observed X-ray binaries.  The X-rays vary over a 35~day cycle;
an $\approx11$ day ``Main-on'' state and $\approx8$ day ``Short-on'' state
(in which the observed X-ray flux is reduced by a factor of $\approx3$)
are separated by half of a 35-day phase (Scott \&\ Leahy 1999).  Outside
of these states the
X-ray flux is several \%\ of that seen in the Main-on state. The X-ray
modulation is not due to a change in the total X-ray output, as the
optical magnitude continues to vary over the 1.7~day orbit due to X-ray
heating of the companion star.  Instead, the 35-day variation probably
results from obscuration of the central source by an accretion disk which
wobbles over a 35-day period due to an unknown cause. X-ray absorption
dips occur at a period of 1.65~days, near to, but significantly greater
than, the 1.62~day beat period between the 1.7 and 35 day periods (Crosa
\&\ Boynton 1981; Scott \&\ Leahy 1999).  Features in the X-ray spectrum
near $30-40$~keV have been interpreted as the result of cyclotron
absorption (Tr\"umper et al. 1978, Dal Fiume et al. 1998), implying a
magnetic field strength $B=2.9\pm0.3\times 10^{12}$~G (Mihara et al.
1990).  Recently, another model has been put forward for the cyclotron
lines, in which the electron momentum distribution is allowed to be
anisotropic. The magnetic field strength is then inferred to be
$B=4-6\times10^{10}$~G (Baushev \&\ Bisnovatyi-Kogan 1999).

The X-ray power spectrum is dominated by the 1.24 second pulse and its
harmonics and by a continuum noise component (Belloni, Hasinger, \&\
Kahabka 1991).  There has been a single report of 144~sec period X-ray
fluctuations during an X-ray turn-on (Leahy et al. 1992). The optical and
UV continuua both show evidence for reprocessed pulsations (Middleditch
\&\ Nelson 1977, Boroson et al. 1996) with fractional amplitudes of
several $\times10^{-3}$.  The accretion disk and the atmosphere of HZ~Her
may both contribute to the reprocessed pulsations.

Quasiperiodic oscillations (QPOs), observed in a variety of compact
accreting binary systems, have not previously been reported for Her~X-1.
QPOs in general are poorly understood.  In the ``atoll'' and
``Z-class'' sources that show kHz QPOs, spin periods are not observed
directly, cyclotron lines are not seen, and emission from the companion
star is often overwhelmed by the emission from the accretion disk. Thus
the discovery of QPOs in Her~X-1 may help to link QPO behavior to
such properties as the neutron star magnetic field and spin, and details
of the mass transfer.

\section{Observations}

We have been carrying out a multiwavelength study of Hercules~X-1
using the HST~Space Telescope Imaging Spectrograph (STIS), the
Rossi X-ray Timing Explorer (RXTE), the Extreme Ultraviolet Explorer
(EUVE), and ground-based observatories, including the Keck Telescope.
Empirical models for the emission lines observed during the first segment
of our campaign, in July of 1998, have been reported in Boroson et
al.
(2000).  A second segment of the campaign took place during July of 1999.
The 1999 observations took place during an ``anomalous low'' state in
which the X-ray flux was two orders of magnitude lower than in the
expected ``main-on'' state, although accretion continued to take place
(Parmar et al. 1999).  A summary of the multiwavelength campaign results
is presented in Vrtilek et al. (2000).


\vbox{
\scriptsize
\begin{center}
{\sc TABLE 1\\
The STIS observation log}
\vskip 4pt
\begin{tabular}{cccccc}
\hline
\hline
Root name & Start & Time & Orbital 
& Count Rate & QPO$\rm ^b$\\
          & (MJD)     & (s) & Phase$\rm ^a$& (s$^{-1}$) & \\
\hline
O4V405010 & 51371.124 & 2227 & 0.512 & 624 & Y\\ 
O4V405020 & 51371.185 & 2636 & 0.549 & 605 & Y\\
O4V405030 & 51371.253 & 2636 & 0.589 & 550 & Y\\
O4V405040 & 51371.321 & 2636 & 0.629 & 491 & Y\\
O4V405050 & 51371.388 & 2620 & 0.669 & 391 & Y\\
O4V406010 & 51372.064 & 2227 & 0.065 & 24 & N\\
O4V406020 & 51372.124 & 2636 & 0.102 & 57 & N\\
O4V406030 & 51372.193 & 2636 & 0.142 & 70 & N\\
O4V406040 & 51372.260 & 2636 &  0.182 & 92 & N\\
O4V406050 & 51372.328 & 2620 &  0.221 & 131 & N\\
O4V407010 & 51373.004 & 2227 &  0.618 & 561 & Y\\
O4V407020 & 51373.065 & 2636 & 0.655 & 435 & Y\\
O4V407030 & 51373.133 & 2636 & 0.695 & 329 & Y\\
O4V407040 & 51373.200 & 2636 &  0.735 & 236 & ?\\
O4V407050 & 51373.268 & 2620 & 0.774 & 158 & ?\\
O4V408010 & 51374.011 & 2227 &  0.210 & 74 & N\\
O4V408020 & 51374.072 & 2636 & 0.247 & 128 & N\\
O4V408030 & 51384.140 & 2636 & 0.288 & 214 & S\\
O4V408040 & 51374.207 & 2620 & 0.327 & 325 & S\\
\hline
\vspace*{0.02in}
\end{tabular}
\end{center}
$\rm ^a${The orbital phase of the mid-exposure time, using the
ephemeris of Deeter et al. (1991)}\\
$\rm ^b${Y=QPOs detecteed, N=Definitely no QPOs detected,
?=Borderline significance, 
S=significant if a linear fit to the light curve is subtracted }
\normalsize
}
\vspace{0.02in}

The UV observations used the HST STIS. The STIS instrument design is
described by Woodgate et al. (1998), and the in-orbit performance of the
STIS is described by Kimble et al. (1998).
We show a log of the 1999 STIS observations in Table~1.
All of these observations used the E140M grating for high
resolution echelle spectroscopy.  This provides a spectral coverage of
$1150-1710$\AA\ with a resolving power of $R=45,800$ (6~\kms).  

\section{Detection of QPOs}

We obtained the HST spectra using the TIME-TAG mode, which stamps each
photon with a time accurate to 125$\mu$sec and the position of the photon
along the two axes that define the two-dimensional echelle image. To
create light curves we first selected a region of the two-dimensional raw
spectral image corresponding either to the entire spectrum or the
brightest spectral lines (\ion{N}{5}$\lambda\lambda1238.8,1242.8$ and
\ion{C}{4}$\lambda\lambda 1548.2,1550.7$). We selected the region in the
two-dimensional echelle image using the task ``fselect'' in the FITS
utility package FTOOLS (version 4.2). Then we converted the resulting file
to a light curve with 0.1~second bins using the FTOOLS ``fcurve'' task.  
We then formed power spectra from the discrete Fourier transform of the
data in each of the HST orbits, and for 400~second intervals within each
orbit.

As a ``control'' on our method, we also examined STIS TIME-TAG data for
the X-ray binary LMC~X-4 using similar data reduction techniques. Analysis
of these observations, and the search for UV pulsations with the
13.5~second LMC~X-4 pulsar period, will be presented in Kaper et al.
(2000).  None of the 5~HST orbits in which LMC~X-4 was observed showed
significant power above the white noise due to counting
statistics in the 1-100 mHz region.

We did not find any evidence for UV pulsations near the 1.24~s pulsar
period in the 1999 data.  We would have detected (with 5$\sigma$
significance) pulsations of rms amplitude $0.2$\%\ near $\phi=0.5$,
amplitude $\approx0.3$\%\ near $\phi=0.75$, and amplitude $\approx0.4$\%\
near $\phi=0.2$ when the disk is emerging from eclipse.  These values
should be compared with the UV rms amplitudes found with the GHRS in
Boroson et al.  (1996): $0.4$\%\ at $\phi=0.56$ and $0.7$\%\ at
$\phi=0.83$. In the 1998 data, during the short-on state at $\phi=0.76$,
we found marginal evidence (false alarm probability $0.4$\%) for
pulsations with rms amplitude of $0.7$\%.  With this sole exception, there
were no UV oscillations near the pulse period in the data from either 1998
or 1999.  The fractional rms amplitude is defined and related to the power
spectrum in van der Klis (1989).  For a constant plus pure sinusoid
signal, the fractional rms amplitude is $1/2 \sqrt 2$ times the
peak-to-peak difference, divided by the average value.

It is useful to place the non-detection of UV pulsations in context by
examining the X-ray pulsations seen in 1998 and 1999.  Our 1998
multiwavelength campaign overlapped with X-ray observations using Bepposax
(Oosterbroek et al. 2000).  These observations found that the fractional
rms amplitude of the X-ray pulses during the July 1998 short-on state was
10\%\ in the 4-10~keV band, compared with 20\%\ in the main-on state (Fig.
4 of Oosterbroek et al. 2000).  RXTE measured an X-ray flux of
$1.5\times10^{-9}$~erg~s$^{-1}$~cm$^{-2}$, steadily decreasing to
$5\times10^{-10}$~erg~s$^{-1}$~cm$^{-2}$ as the short-on state progressed.
The 1999 RXTE observations found that
although the direct component of the X-rays was obscured, X-rays were
observed at $\approx5$\%\ of the normal main-on level, and
continued to pulsate
with the 1.24~second period, with a fractional rms amplitude of
$1-3$\% (Still et al. 2000).  We caution, however, that these low pulse
fractions
do not provide a full explanation of why UV pulses were not observed,
as the X-ray pulse fraction seen by the companion star and disk (which
cause the UV pulsations) can
be very different from that which we observe.

\vbox{
\vspace{0.4in}
\epsscale{0.8}
\plotone{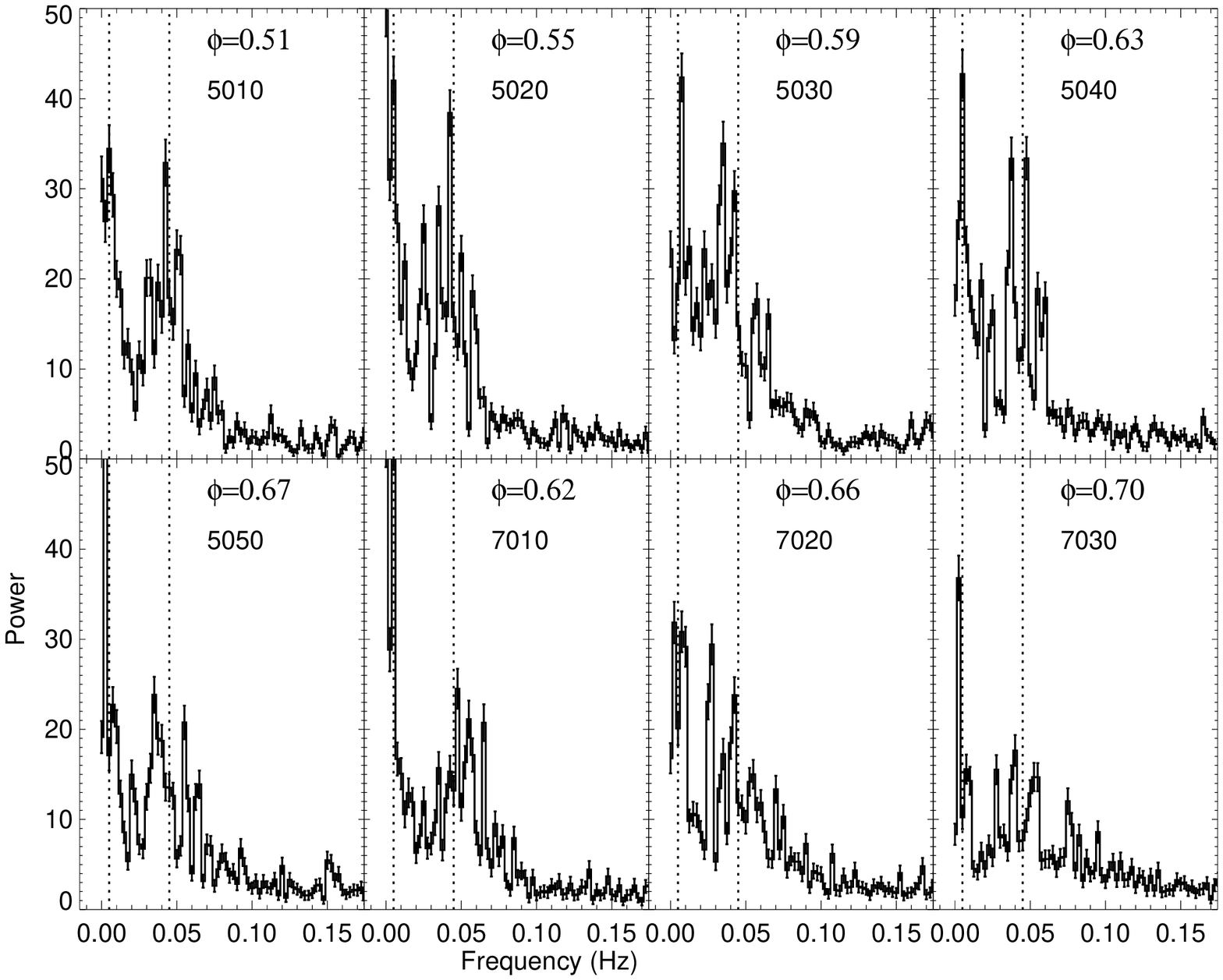}
\vspace{0.5in}
\figcaption{Individual power spectra for the continuum light curves of
each HST orbit in which QPOs were detected.  We label the mean orbital
phase and the observation root name for each orbit.  The dotted vertical
lines mark 5 and 45 mHz.}
}

\vspace*{0.2in}

In Figure~1 we show the average power spectrum for each orbit in which
QPOs were detected.  In Figure~2 we show the average of 46~power spectra
of the light curves from all HST orbits from the 1999 campaign in
400~second intervals.  Near $\phi=0.5$ the count rate is sufficient to
detect individual QPO oscillations, which we show in Figure~3.  No
significant QPOs were detected in the emission line flux. The lines
contribute about 20\%\ of the total counts in the spectrum. When we
subtract the light curve of the emission lines from the light curve of the
total spectrum and take a power spectrum, the QPOs remain unaffected.

\vbox{
\epsscale{1.15}
\plotone{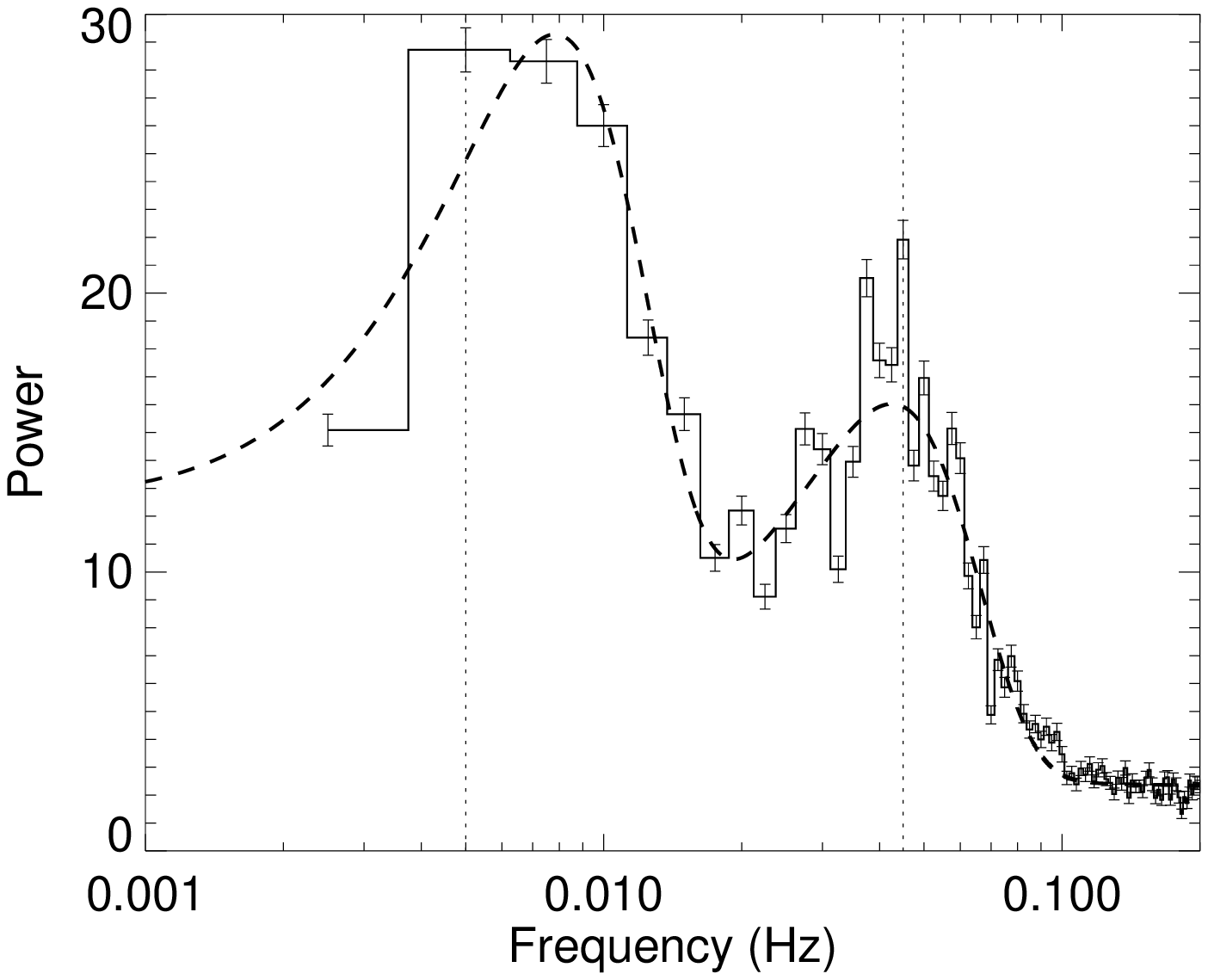}
\figcaption{The sum of 46 power spectra of 400~second light curves of the
UV continuum of Her~X-1
observed in July 1999 with the STIS aboard HST. The noise level is
normalized to 2 (Leahy normalization).  The dashed curve shows a Gaussian
fit to 
the two peaks.  The dotted vertical lines mark 5 and 45 mHz.)}
}

\vspace*{0.2in}

We could not obtain accurate Gaussian representations of the power spectra
within each HST orbit.  However, taking the average of all power spectra,
we could fit a power-law continuum and 2~Gaussians (Figure~2).  We find
QPO centroids of $\nu_1=8\pm2$~mHz and $\nu_2=43\pm2$~mHz (we have
determined errors from a Monte Carlo bootstrap analysis, Press et al.
1992).  The gaussian widths of the QPOs were $4\pm1$~mHz and
$21\pm2$~mHz. Applying
the techniques outlined in van der Klis (1989), we find
total rms QPO fractions of 2.1$\pm0.7$\%\ and 3.7$\pm0.4$\%\ for the two
QPO peaks, respectively.  

We detect QPOs at $\phi=0.28-0.33$ if we remove a linear trend from the
UV continuum light curve (Table~1).  It is possible that the
predominance of UV oscillations near $\phi=0.5$ results from the flatter
orbital light curve near $\phi=0.5$.  However, removing linear trends from
the data for other HST orbits did not reveal mHz~QPOs.

After finding QPOs in the 1999 STIS data, we examined optical observations
from our 1998 campaign and found evidence for a 35~mHz QPO peak there as
well (Figure~4).
Results from the optical campaign will be presented in more detail in
(O'Brien et al. 2000).
The HST~GHRS
observations of Boroson et al. (1996) during the Main-On state included
1200~s of high count-rate (1100~cts~s$^{-1}$) data at $\phi=0.55$, which
show the 8~mHz QPO but not the 43~mHz QPO.  

\vbox{
\epsscale{1.15}
\plotone{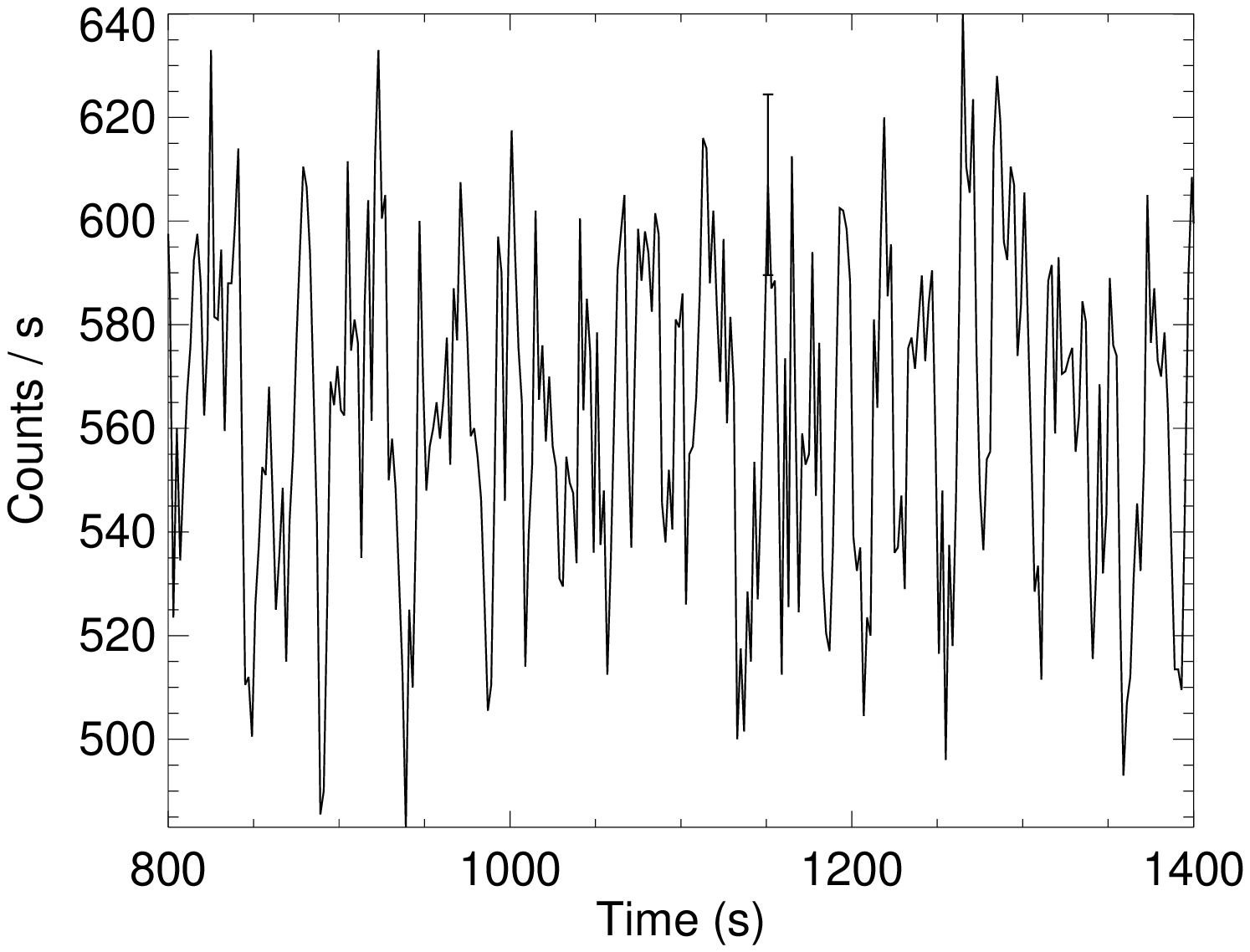}
\figcaption{A portion of the UV continuum light curve
showing individual QPO oscillations.  The x-axis shows the time elapsed 
from the start of observation 5030.  A single error bar is shown; errors 
on the other points are similar.  Data are presented in 2~second
time bins.  The rms variability for this portion of the data is 5.6\%,
while the expected noise rms is 3.0\%.  This is consistent with a QPO rms 
of 4.7\%.}
}


\vbox{
\epsscale{0.66}
\plotone{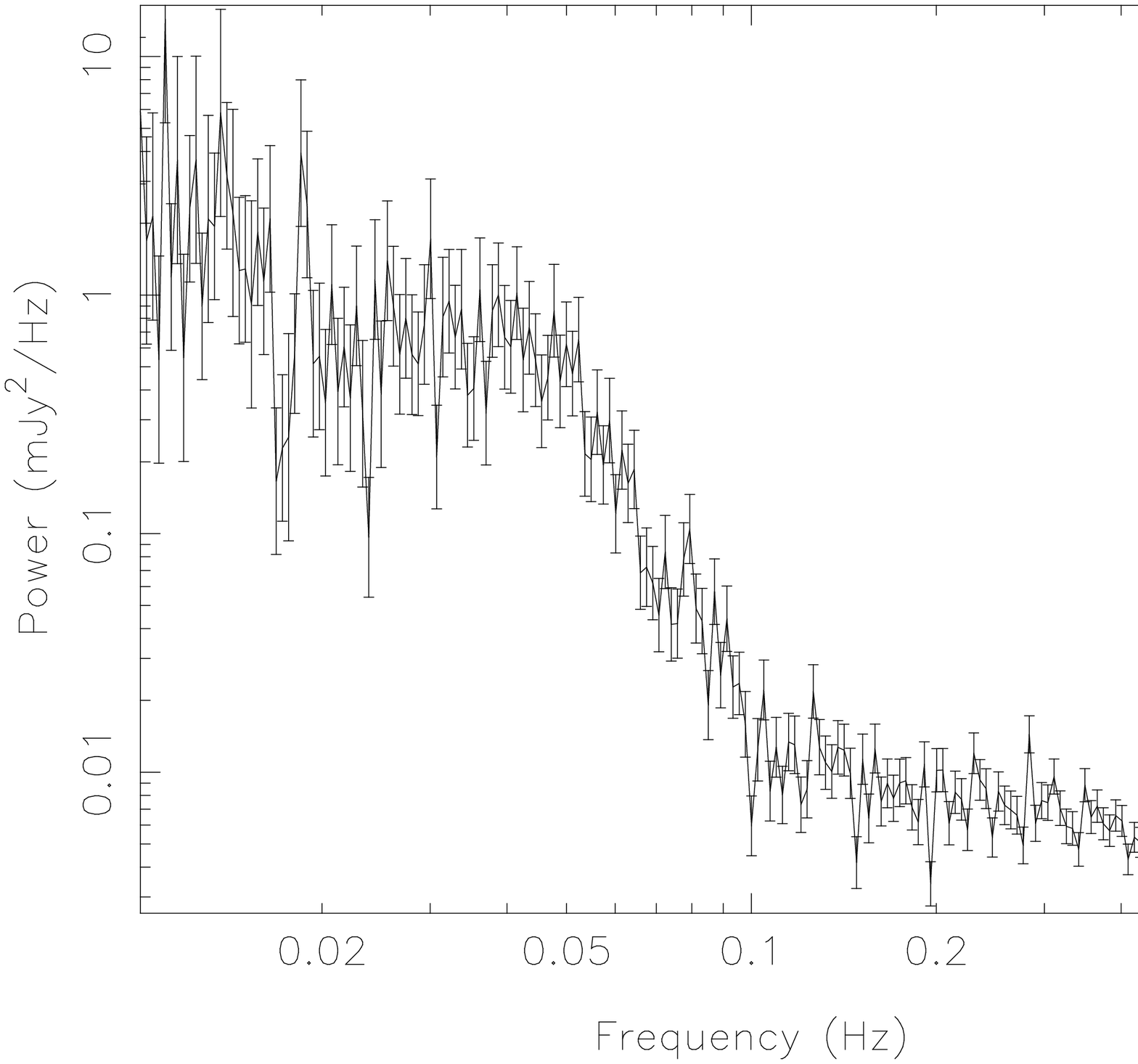}
\figcaption{Power spectrum of the optical continuum of Her~X-1 observed
in
July 1998 with the Keck Telescope.}
}
\vspace{0.1in}

We investigated the time dependence of the 45 mHz QPO feature using a
Gabor transform (Heil \&\ Walnut 1989, 1990).  The Gabor transform is a
wavelet transform which decomposes the signal into the time-frequency
plane. The discrete implementation used here is equivalent to a
short-time, Gaussian-windowed Fourier transform (Boyd et al. 1995).  We
chose a window size of 512 seconds, and restricted the frequency range to
lie between 2.5 - 90 mHz.  Figure~\ref{fig:gt} displays power as a
function of frequency and time for the first five observations.  The
palette runs from dark blue (low power) through deep red (high power).  
The featureless dark blue vertical strips are times during which no data
was
collected.  Significance was estimated by performing individual,
independent Fourier transforms on the data in windows fully containing
the features of highest power.  The strongest of these features are found
to be significant at more than $\sim6\sigma$.

Figure~\ref{fig:gt} shows the frequency evolution of the 45~mHz QPO as a
function of time, and can be thought of as an
unfolding
of the single power spectrum, which shows a broad QPO feature centered
about a preferred frequency.  The frequency evolution shown in the Gabor
transforms shows 
that the QPO in the overall power spectrum
is composed of many individual short-time components that come and go
over the entire observation. 

\vbox{
\epsscale{0.65}
\plotone{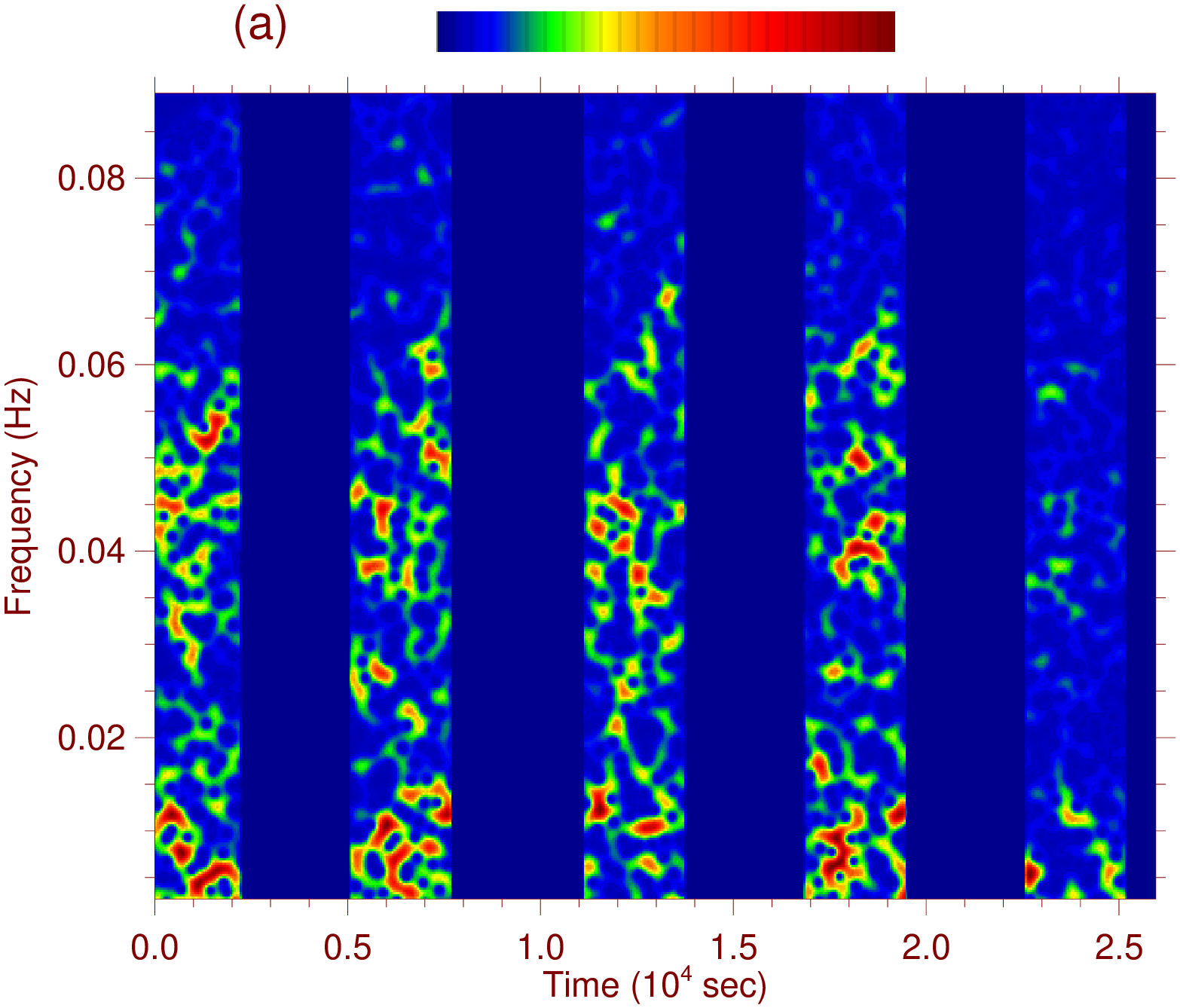}\\
\vspace{0.75in}
\epsscale{0.65}
\plotone{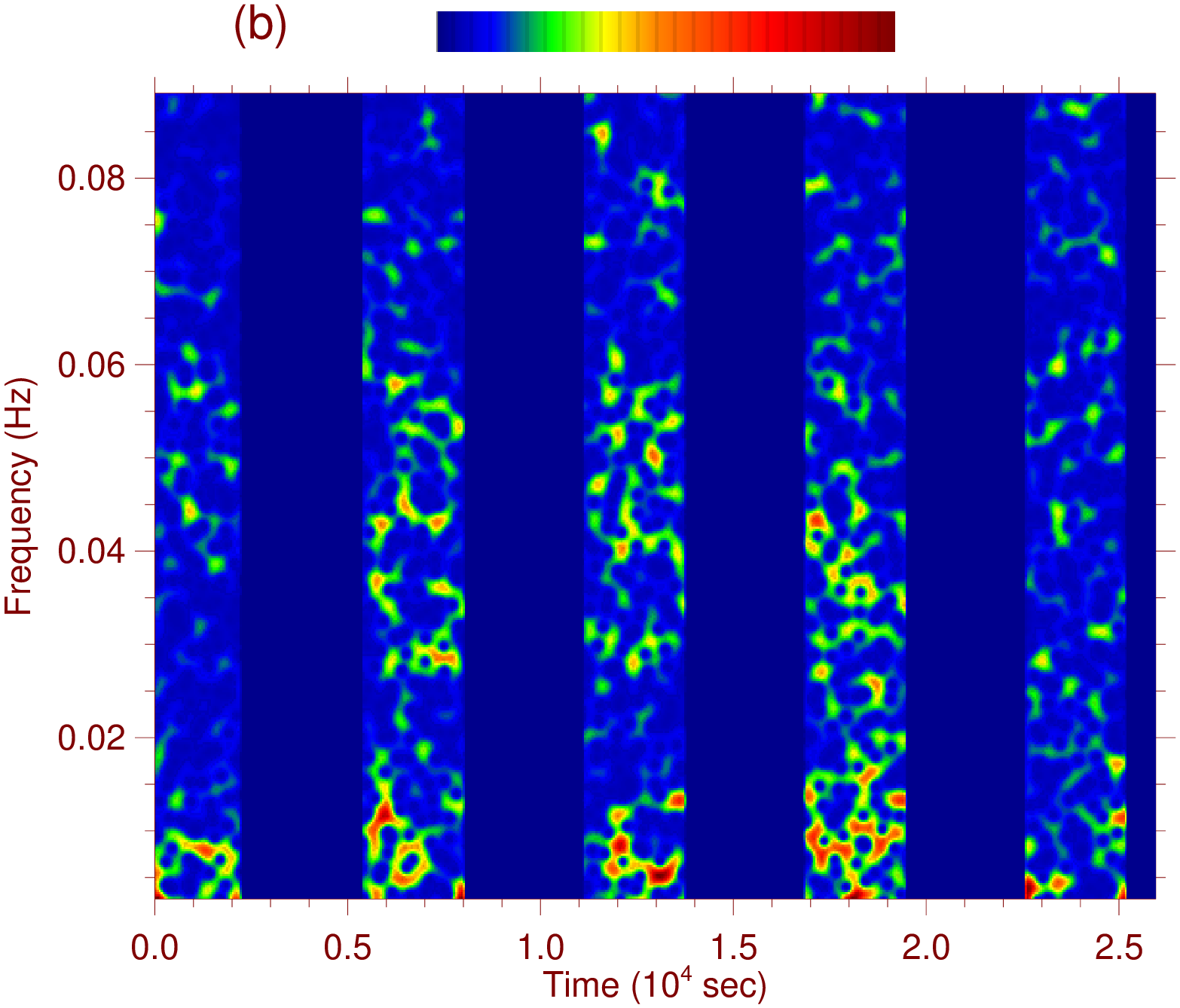}
\vspace{1.0in}
\figcaption{\label{fig:gt} The Gabor Transform, a dynamical power
spectrum, applied
to the observations for which QPOs were detected.  Colors range from
dark blue (low power) through red (high power) according to the 
palette shown.  The frequency range covers both mHz QPO features.
Neither feature is constant in time, but made up
of shorter, apparently more coherent, signals.  Features with the
highest power are significant at $\approx6\sigma$.
There is no obvious correlation between the instantaneous power or
frequency of one feature with the other.  (Dark blue vertical bands
represent times when no data was obtained.) Panel (a) shows the 50x0
series of observations,
and (b) shows the 70x0 series.}
}


\section{Discussion}


The ``spikey'' power spectra could be caused by several processes.
Multiple overlapping exponential oscillating shots or a single sinusoid
whose phase undergoes a random walk could each reproduce the qualitative
power spectral shape observed.  The Gabor transforms
suggest that the QPO width is due to high Q oscillations appearing and
disappearing at different frequencies.
Oscillating shot noise can cause both QPO
and a low-frequency red noise component together, and we have tried to
determine whether the 8~mHz feature could actually be red noise
associated
with the 45~mHz feature.  Analysis of the dynamical power spectrum through
the Gabor transform shows no apparent correlation.

Are the QPOs due to the accretion disk or the X-ray heated atmosphere of
HZ~Her? The QPOs were not detected at $\phi\lae0.33$ or $\phi\gae0.73$.  
Models suggest that at $\phi<0.2$ and $\phi>0.8$, most of the optical and
UV continuum arises in the accretion disk, and that from $\phi=0.2$ to
$\phi=0.8$ the flux from the disk should be constant (Howarth \&\ Wilson
1983).  The X-ray heating of the atmosphere of HZ~Her causes the optical
continuum flux to reach a maximum near $\phi=0.5$ (Bahcall \&\ Bahcall
1972), and at these phases, the UV emission arises predominantly on the
X-ray heated face HZ~Her (Vrtilek \&\ Cheng 1996).  From Figure~6 we
conclude that at $\phi=0.5$, the disk continuum should contribute at most
$10$\%\ of the total continuum flux.  Thus if the UV QPOs with 10\%\
amplitude of the total flux result from oscillations of the disk flux, the
disk flux must be modulated by an amplitude near 100\%.

\vbox{
\epsscale{1.}
\plotone{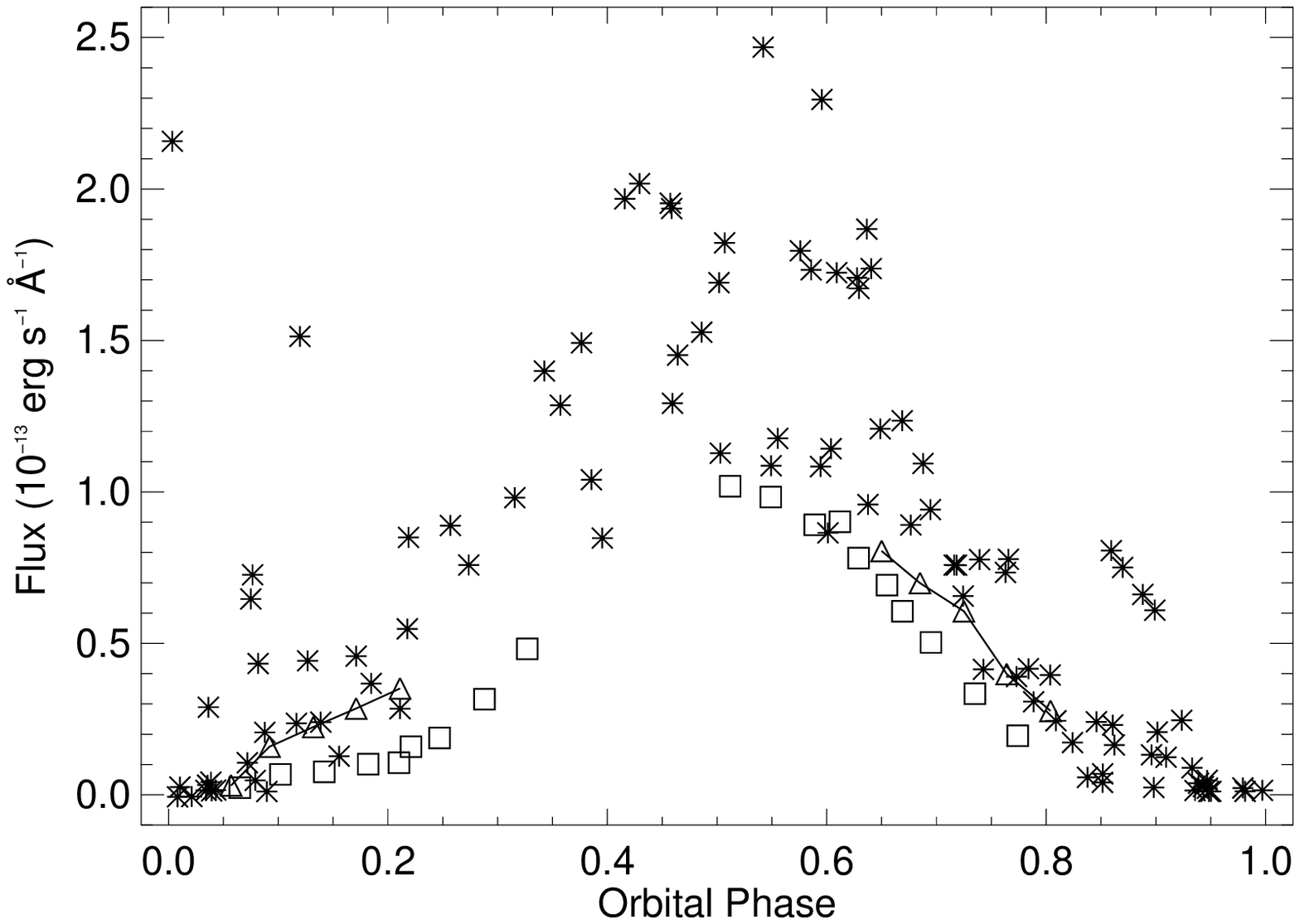}
\figcaption{Flux of the UV continuum (1260\AA\ to 1630\AA) as
observed with IUE, versus
orbital phase.  The diamonds show the fluxes observed with the HST STIS
in July 1998, and the squares show the fluxes observed in July 1999.}
}

\vspace*{0.2in}

It would be simpler to attribute the QPOs to the heated atmosphere of
HZ~Her.  The concentration of the QPO detections about $\phi=0.5$ could be
a result of the increased stellar flux; with a higher flux the QPOs would
be easier to detect. The light-travel time across the X-ray heated face of
the companion should be $<10$~seconds, as the separation of the neutron
star and the system center of mass is $a \sin i=13.186$~s (Deeter et al.
1991).  This allows variability on the observed timescales, while damping
the harmonics of the 45~mHz QPO (which are not observed).

\vbox{
\scriptsize
\begin{center}
{\sc TABLE 2\\
X-ray Binary Pulsars with mHz QPOs}
\vskip 4pt
\begin{tabular}{lcccccl}
\hline
\hline
Name & Type$\rm ^a$ & $\nu_{\rm pulse}$ & $\nu_{\rm QPO}$ & B &
Reference\\
     &      &  (mHz) & (mHz) & $10^{12}$~G & \\
\hline
Her X-1 & P & 807.9 & $5,45$ & 2.9 & this
work\\ 
        &   &       &        & $\pm0.3$ & \\
SMC X-1 & P & 1410 & $10,60$ &  & Wojdowski\\
        &   &      &         &  &  et al. 1998\\
4U 1627-67 & P & 132 & 48 & $\approx3$ &  Chakrabarty\\
       &          &      &  &  &  1998\\
Cen X-3 & P & 207 & 35 & &  Takeshima \\
        &   &     &    & &  et al. 1994\\
4U 0115+63 & T & 277 & 62 & 1.0 &  Soong \&\ \\
           &   &      &   &     &  Swank 1989\\
4U 1907+09 & T  & 2.27 & 55 &     & in'tZand \\
           &      &    &    &    & et al. 1998\\
XTEJ1858+034 & T & 4.5 & 110 &  & Paul \&\ Rao \\
              &   &    &     &   & 1998\\
V0332+53 & T & 229 & 51$\pm5$ &  &  Takeshima \\
         &   &     &          &  & et al. 1994\\
EXO 2030+375 & T & 24 & 213 &  & Angelini et al.\\
              &    &   &    & &  1989\\
A 0535+262 & T & 9.71 & 25-72 & & Finger et al.\\
                  &   &    &     & &  1996\\
\hline
\vspace*{0.02in}
\end{tabular}
$\rm ^a${P: persistent X-ray source, T: transient X-ray source}\\
\end{center}
\normalsize
}

Although we conclude that the UV QPOs are probably emitted by HZ~Her, they
may well be {\it reprocessed} from QPOs originating in the accretion disk,
either as a result of some structure in the disk {\it emitting} with the
QPO frequency, or {\it shadowing} the star from X-ray emission with the
same frequency.  A comparison with optical and X-ray QPOs recently
discovered in other X-ray pulsars supports this conclusion (see Table~2
for a summary of these systems.)

The KZ~TrA/4U~1627-67 system contains an X-ray pulsar with a 7.7~s period,
and shows 48 mHz QPOs in both X-ray and optical lightcurves (Chakrabarty
1998).  The optical QPOs, with an amplitude of $3-5$\%, are emitted by the
surface of the companion in response to X-ray illumination.  The QPO
frequency is probably the orbital frequency of a structure revolving about
the neutron star (Kommers, Chakrabarty, \&\ Lewin 1998).  The X-ray binary
pulsar SMC~X-1
shows X-ray power spectral turnover at $\sim10$~mHz (Angelini, Stella,
\&\ White 1991) and QPOs at 60~mHz (Wojdowski et al. 1998).  Transient
X-ray pulsars also show QPOs; XTE J1858+034 has QPOs with $\nu=110$~mHz
with rms amplitude 3.7-7.8\%\ depending on the energy band (Paul \&\ Rao
1998). The transient
Be system 4U~1907+09 shows very narrow QPO with $\nu\approx55$~mHz during
bursts (in'tZand, Baykal, \&\ Strohmayer 1998).

These sources thus show QPOs with remarkably similar frequencies, and as
they are all X-ray pulsars they probably have similar magnetic field
strengths (for 4U~1627-67, BeppoSAX observations of a cyclotron feature
imply $B\approx3\times10^{12}$~G, Orlandini et al. 1998). It thus seems
natural to try to associate the QPOs with the region where the magnetic
field disrupts the disk.  The Alfv\'en radius (where the magnetic energy
density equals the gas ram pressure) in Her~X-1 is probably
$r_A=2-7\times10^{8}$~cm (McCray \&\ Lamb 1976). Within this region, the
Keplerian frequency is $>100$~mHz, higher than the observed QPOs.  
The QPO may be due to Keplerian rotation further out in the disk,
but a physical reason why this region is singled out is lacking.

The 8 mHz QPO in Her~X-1 is even more problematic.  If it occurs as the
result of Keplerian rotation, it arises at a radius in the disk of
$r=6\times10^{9}$~cm, even further from the Alfv\'en radius.

We also consider a ``beat-frequency'' interpretation (Alpar \&\ Shaham
1985), in which
the observed QPO frequency $\nu_{\rm QPO}=45\mbox{mHz}=\nu_{\rm ns}-
\nu_K$, where $\nu_{\rm ns}=808\mbox{mHz}$ is the neutron star
spin frequency and $\nu_K$ is the Keplerian frequency of some material
in the disk.  In order for $\nu_{\rm QPO}$ to be a beat frequency,
the Keplerian frequency must be within 5\%\ of $\nu_{\rm ns}$.
This seems an unlikely coincidence, although we must have
$\nu_{\rm ns}<\nu_K$ or the gas will be expelled from the disk, spinning
down the pulsar via ``propellor effect'' (Illarionov \&\ Sunyaev 1975).
We note that Bepposax observations ending $<2$~days before the start
of the HST observations reported here (Parmar et al. 1999) show
that the pulsar has indeed spun-down, reversing $\approx5$ years of
spin-up in a period of months.  The beat frequency interpretation
of the QPOs requires that the Alfv\'en radius be near the corotation
radius in Her~X-1 (at $2\times10^8$~cm), and this does appear to be the
case. 

The transition between Keplerian rotation and corotation occurs at a
radius $r_0\approx0.5 r_A$ (Ghosh \&\ Lamb 1979) over a radial zone of
extent $\Delta r=0.03r_0$.  In this range, $\nu_K$ varies by
$\approx5$\%, naturally leading to $\nu_{QPO}\approx50$~mHz, given
$\nu_K=\nu(r_0)\approx\nu_{\rm ns}$.  (However, this mechanism
would not be able to explain the 8~mHz QPO, which requires the
Keplerian frequency to be within $\approx0.5$\%\ of the pulse
frequency.)  

Further bolstering the beat-frequency case, we have found marginal
evidence for oscillations at 760~mHz=$808-45$~mHz in the X-ray data from
RXTE from our 1998 campaign.  As an example of how this could arise in a
beat-frequency scenario, if a blob in orbit around the neutron star has an
orbital frequency of 760~mHz, it will cover the X-ray beam from the pulsar
(which rotates at 808~mHz) and prevent it from illuminating the companion
star with a frequency of $808-760$~mHz$=48$~mHz.  The (marginally)
observed 760~mHz X-ray signal would then be a direct signal from this
blob, due either to emission or obscuration.

It seems more likely that the reprocessed QPOs are due to material that
absorbs the X-rays and prevents them from heating HZ~Her than material
that emits X-rays.  Whether the QPO frequency results from Keplerian 
rotation or a beat-frequency, the accretion disk is too cold at the
implied radii to emit X-rays that significantly heat the companion.

A third possible explanation for the QPOs, that may explain the presence
of {\bf both} QPOs, is provided by Titarchuk \&\ Osherovich (2000,
submitted).  In this explanation, the higher frequency QPO results from a
global vertical oscillation of the disk, whereas the lower frequency QPO
results from Keplerian rotation.

We detected UV continuum QPOs during an anomalous low state of Her X-1.
Although there may be a connection between the QPO and the low state,
time-resolved UV observations near $\phi=0.5$ have not been common, and
similar optical QPO are seen in observations using the Keck telescope
(O'Brien, et al. 2000) during the short-on state (Figure~4). The IUE
observations and
the HST~FOS observations of Anderson et al. (1994) did not have the
required time resolution for the detection of QPOs. 

The QPO phenomenon in Her~X-1 should help us understand the origin of QPOs
in other sources.  The simultaneous detection of X-ray and
UV QPOs during an On state could test our suggestion that HZ~Her
reprocesses X-ray QPOs.  
The UV QPOs should lag the X-ray QPOs by $\approx10$~s, in a manner
that varies with the orbital phase.
Simultaneous measurement of UV and optical QPOs could determine the
spectrum of the reprocessor. These analyses will test our proposal that
the QPOs originate on the star in response to disk QPOs, and may help us
distinguish between the possible origins in the disk that we have
suggested.

\acknowledgements

We would like to thank Vladimir Osherovich and Lev Titarchuk for
discussions and suggestions.
Based on observations with the NASA/ESA {\it Hubble Space Telescope},
obtained at the Space Telescope Science Institute, which is operated by
the Association of Universities for Research in Astronomy, Inc., under
NASA contract GO-05874.01-94A.  BB and SDV supported in part by NASA
(NAG5-2532, NAGW-2685), and NSF (DGE-9350074).  BB acknowledges an NRC
postdoctoral associateship.  HQ is employed on PPARC grant L64621.
We would like to thank R. Kelley and D. Chakrabarty for suggestions.

\end{document}